# An Algorithm for Recommending Groceries Based on an Item Ranking Method

Gourab Nath
Department of Data Science
Praxis Business School
Bengaluru, INDIA
email: gourab@praxis.ac.in

Jaydip Sen
Department of Data Science
Praxis Business School
Kolkata, INDIA
email: jaydip@praxis.ac.in

*Abstract*— This research proposes a new recommender system algorithm for online grocery shopping. The algorithm is based on the perspective that, since the grocery items are usually bought in bulk, a grocery recommender system should be capable of recommending the items in bulk. The algorithm figures out the possible dishes a user may cook based on the items added to the basket and recommends the ingredients accordingly. Our algorithm does not depend on the user ratings. Customers usually do not have the patience to rate the groceries they purchase. Therefore, algorithms that are not dependent on user ratings need to be designed. Instead of using a brute force search, this algorithm limits the search space to a set of only a few probably food categories. Each food category consists of several food subcategories. For example, "fried rice" and "biryani" are food subcategories that belong to the food category "rice". For each food category, items are ranked according to how well they can differentiate a food subcategory. To each food subcategory in the activated search space, this algorithm attaches a score. The score is calculated based on the rank of the items added to the basket. Once the score exceeds a threshold value, its corresponding subcategory gets activated. The algorithm then uses a basket-to-recipe similarity measure to identify the best recipe matches within the activated subcategories only. This reduces the search space to a great extent. We may argue that this algorithm is similar to the content-based recommender system in some sense, but it does not suffer from the limitations like limited content, over-specialization, or the new user problem.

*Keywords—item recommendation, grocery item recommender system, item ranking, basket-to-recipe similarity, bulk recommendation.*

## I. INTRODUCTION

A large number of buyers come online to do their grocery shopping. And let's say there waits an algorithm that watches every buyer as they shop. Based on what they add in their baskets, the algorithm attempts to figure out the possible things they can (or intend to) cook. On recognizing those, the algorithm steps forward as a bot and presents a list of suggestions to the users. As the user chooses one or more options from the list, it helps the buyers to add the rest of the required ingredients in their respective baskets. This problem is relatively new in the areas of recommender systems and is the focus of this research.

The presence of such an application can enhance the overall buying experience of the customers. This application can save a lot of buyers' time and also help to prevent item miss outs. Nevertheless, such a new shopping experience may also help to kill the monotony of regular online grocery shopping. In this paper, we introduce such a recommender system for grocery shopping.

Formally, this problem can be formulated as follows. Let $C$ be the set of all food categories. A food category $c \in C$ can be classified into a set of food subcategories $S_c$. Each subcategory $s_c \in S_c$ consists of a set of dishes $D_{sc}$ and each dish $d_{sc} \in D_{sc}$ are prepared using a set of items $I_{dsc}$. For example, the set $C$ may contain categories like *rice, noodles, chicken, fish, etc.* The food subcategory $S_{rice}$ may contain subcategories like *fried rice, biryani, pulao, etc.* And $D_{fried\ rice, rice}$ may contain dishes like *chicken fried rice, egg fried rice, Singapore fried rice, Szechuan fried rice, etc.,* and the set of items which are used to prepare these dishes forms the set $I_{dsc}$.

This work is organized as follows. Section II provides a discussion on what makes this algorithm different for grocery item recommendations while discussing some of the related works. Section III provides a detailed methodology of the proposed algorithm. Section IV provides a discussion on the kind of data to be used and some detailed results. Finally, Section V concludes the paper.

## II. RELATED WORKS

Recommender systems are usually classified into three categories, viz. content-based recommender system, collaborative recommender system, and hybrid recommender system [1]. A Content-based recommender system recommends those items to a user that are very similar to the items previously preferred by the user [2, 3]. The preference or the usefulness of an item to the user can be measured using a utility function [4]. The utility of an item to a user is usually represented by a rating given by the user to express his/her item preference. In a content-based recommender system, the similarity between any two items is calculated based on the assumption that there exists a set of features that are explicitly associated with the items to be recommended [5, 6].

Generating a set of explicit features that can sufficiently represent the grocery products may be an extremely tedious task to do. Manual assignments of these features to the products are often not practical due to the limitation of resources [7]. Moreover, users may not have the patience to rate all the products they are purchasing. Also, a poor rating given by a user to an item need not necessarily mean that such items are not very useful to the user. For example, if a user has rated some fruits or vegetables poorly, then that may be because they were not delivered fresh to the user and probably not be because the user did not like to consume the products. Since our algorithm is not dependent on user rating or the product contents, it does not suffer from these problems.

A quality recommender system should recommend the kind of items that a user may prefer but has never seen before. Our algorithm is capable of making such recommendations [8, 9]. For example, a user who picks up a packet of *chips* and *mayonnaise* may get tempted to pick up something which he/she has never purchased before, like say, a jar of *schezwan sauce* and a packet of *grated mozzarella cheese* after getting to know that he/she could prepare an easy a beautiful cheesy snack to spice up his/her evening. Also, it is better to avoid recommending items that are too similar [10]. For example, if a user highly rates a packet of *wheat* of a particular brand, then it doesn't make much sense to recommend that user *wheat* of several other brands. Content-based filtering algorithms suffer from this problem of over-specialization [11, 12]. However, our proposed algorithm does not suffer from such limitations and problems.

A collaborative recommender system [13, 14, 15] attempts to recommend items to a user that is preferred by some other users who are similar to that user. Similarities of two users are calculated based on how their ratings on different items. This algorithm heavily depends on the ratings given by the users. The other advanced algorithms discussed in [16] also depend on the user ratings and will face the same challenge discussed above. We must also point out that our algorithm does not suffer from the cold start problem [11,17, 18].

## III. METHODOLOGY

One simplistic framework for solving this problem could be as follows. Let $B$ denote the set of items added to the basket by a customer. Initially, $B$ is an empty set. Calculate the cardinal number of the set $I_{dsc} - B$, for all $d, s$ and $c$, every time a new item is added to B. The set $I_{dsc} - B$ would contain all the items that are present in the item set of the dish $d$ but is not present in basket B. The value of $n(I_{dsc} - B)$ can be seen as a measure of closeness of the basket $B$ from the dish $d$. From here on, we will call this metric the *basket-to-recipe similarity*. If $n(I_{dsc} - B) < \theta$, a certain threshold value, then we may recommend the dish $d$ to the customer.

This approach, however, has some potential disadvantages. Firstly, this algorithm doesn't give any special attention to the items which are considered as important for some specific food categories and subcategories. Secondly, $n(I_{dsc})$ is not constant for all $d$. Therefore, the above process may tend to return those dishes for which $n(I_{dsc})$ is small even if the added items may be highly representative of some other dishes for which $n(I_{dsc})$ may be large. Thirdly, every time an item is added to the basket, the algorithm compares $B$ with all $I_{dsc}$ Which seems to be a bit unnecessary. This makes the algorithm unnecessarily complex. The looping process should skip on the addition of the most commonly used items like *salt* and *sugar* in the basket.

In this research, we propose an algorithm that takes care of these problems. We suggest some improvements to this simplistic framework keeping the following two observations in mind. (1) Each food category $c$ can be identified by the set of the most frequently occurring items in that category. We term these items as *category identifiers*. (2) Each subcategory $S_c$ in $c$ can be recognized using a set of special items that help in differentiating $S_c$ from the other subcategories in $c$. We term these items as *subcategory differentiators*. Instead of calculating the basket-to-recipe similarity for every $I_{dsc}$ we limit our search space to only a subset of $I_{dsc}$. A category is activated only when a category identifier is observed, and a subcategory is activated only when a number of subcategory differentiators are observed. The search space is kept limited to the dishes belonging to the activated subcategories only. This means, instead of calculating the basket-to-recipe similarity for all $I_{dsc}$, we limit our calculations to only the activated $I_{dsc}$. This reduces the time complexity of the algorithm to a great extent.

### A. Activating a food category

Each category can be identified using a set of items that occur most frequently in that category. Let $R$ represent the set of unique items used across all the dishes. More formally, $R = \bigcup_c \bigcup_s \bigcup_d I_{dsc}$. For each category $c$, a sparse binary matrix $M_c$ is created such that each row of $M_c$ represents a dish that belongs to the category $c$ and the columns represent the items in $R$. The values of the matrix $m_{c(i,j)}$ take the value 1 if the $j^{th}$ item is present in the recipe of the $i^{th}$ dish. Otherwise, it takes the value 0. The dimension of $M_c$ is $n(\bigcup_s D_{sc})$ by $n(R)$, where $n(.)$ is used to represents the cardinal number of a set.

We use the sparse binary matrix $M_c$ to calculate the support of the items in $I_{dsc}$. The support of an item in a given category $c$ is defined as the probability of observing that item in that category. The items with the top $k$ support values are used as the category identifiers.

However, we must ensure that none of these items have very high global support. Global support of an item is calculated by calculating the probability of occurrence of the item from the data $\bigcup_c M_c$. Items having high global support are common items like salt and sugar. These common items should not be used as item identifiers. More formally, if $F_c = \{f_c^1, f_c^2, \dots, f_c^k\}$ defines the set of category identifiers for the category $c$ and $I = \{I_1, I_2, \dots, I_p\}$ defines the set of items with top $p$ global support values then $F_c$ should be updated as $F_c := F_c - I$. Selection of any $q$ items from $F_c$ activates the category c.

*Algorithm 1. Food category activation algorithm.*

*Comment: Function to find the set of k category identifiers for category c.*

$I_{global} \leftarrow$ items with top $h$ global support values

**function** *category_identifiers* $(c, k)$
    $M_c \leftarrow$ load the matrix $M_c$
    $R \leftarrow$ column index of $M_c$
    **for** item in $R$ **do**

```
    sup ← sum (M_c[ , item]) / row_count(M_c)
    sup_pair [i] ← (sup, item)    {storing as pairs}

  sup_pair ← sort(sup_pair) by sup
  identifier ← empty array
  i = 0
  while length(identifier) ≤ k do
      i = i + 1
      if sup_pair [i][1] is not in I_global then
          add sup_pair [i][1] to identifier
  return(identifier)
```

*Comment: Function to update the activation score on the addition of a new item in B. Here T is an array whose $i^{th}$ element will store the activation score of the $i^{th}$ category*

```
function activation_score (T: array, item)
    C ← set of all categories
    count ← 0
    for c in C do
        count += 1
        if item in category_identifiers (c , k) then
            T [count] += 1
    return (T)
```

*Comment: Process to activate a category. If the addition of an item makes the $i^{th}$ value of T exceed the threshold value k, then the $i^{th}$ category will be activated.*

```
T ← [0, 0, …, 0]             {initialization to zero}
active_cat ← set of active category

while item added in B:
    T ← activation_score (T, item)
    for i from 1 to length(T) do
        if T [i] > k then
            active_cat ← add the i^{th} category to the set
                         active_cat
            print (The category c is activated)
```

### B. Activating a food subcategory

Each subcategory $s_c$ can be distinguished from the other subcategories using a set of few items which are commonly used to prepare the dishes that belong to $s_c$ but rarely used to prepare the dishes that belong to the subcategories other than $s_c$ for a given $c$. To identify the subcategory differentiators of $s_c$, we first find the conditional probability $P(R = r \,|S_c = s_c)$ for all $r \in R$. Note that a higher value of this probability indicates the popularity of the item in that subcategory. However, it should also be noted that popular items need not necessarily be a differentiator. In fact, if an item is very popular in a particular subcategory, then it is likely to be popular in the other subcategories.

Next, for each $r \in R$ we calculate the difference $P(R = r \,|S_c = s) - \sum P(R = r \,|S_c = s_c)$, where the summation is over all values of $S_c$ other than $s$. We use the notation $p_{rs}$ to denote this measure. The upper limit of $p_{rs}$ is 1. A value of $p_{rs} = 1$ indicates that the item $r$ is common to all the dishes in the subcategory $s$ and is not used in any other dishes that belong to the subcategories other than $s$. If an item is a very common item across all the subcategories in $c$ then the value of $p_{rs}$ will be small and can even be negative. After $p_{rs}$ is calculated, we rank each item according to the values of $p_{rs}$ for each $s$. For a given subcategory, the item that corresponds to the maximum value of $p_{rs}$ is ranked 1. The $rank(p_{rs})$ shows the importance of the item $r$ in distinguishing the subcategory $s$ from the other subcategories in $c$.

The addition of a few top-ranked items should activate a subcategory. For each item added to the basket, we use the formula in (1) to score each subcategory $s$ for an active category $c$.

$$score_{sc} = score_{sc} + \frac{1}{\sqrt[n]{rank(p_{rs})}} \quad (1)$$

If the value of $score_{sc}$ exceeds a certain threshold value $\theta$, then we activate the subcategory $s$.

From the above equation, it is evident that the top-ranked products will add more to the score. The value of $n$ controls the number of items required to activate a subcategory. If the value of $n$ is large, then the addition of less number of observations to the basket can activate a subcategory, whereas a small value of $n$ will require more number of observations to activate a subcategory. Fig. 1 shows how the function $\frac{1}{\sqrt[n]{x}}$ varies for different values of $n$.

*Algorithm 2. Food subcategory activation algorithm.*

*Comment: Function to find the $rank(p_{rs})$ for a given category c.*

**function** *item_rank*(c)

*Comment: the matrix H will store the calculating the conditional probabilities $P(R = r \,|S_c)$. Each row of H will represent a subcategory in c, and the columns are items in R.*

```
H ← an empty matrix
i ← 0
j ← 0
for s in S_c do
    i ← i + 1
    for item in R do
        j ← j + 1
        prob ← sum (M_c[ i,  j]) / row_count(M_c)

        H[i, j] ← prob
```

*Comment: The matrix B will store the values of $p_{rs}$. The row labels and column labels of B are the same as that of H. Note, H[-i, j] represents the $j^{th}$ column of H except the $i^{th}$ row*

```
B ← an empty matrix
for i from 1 to row_count(H) do
    for j from 1 to column_count(H) do
        p = H[i , j] - sum(H[-i, j])
        B [i, j] ← p
```

*Comment: The matrix Rank will store the values of $rank(p_{rs})$. The row labels and column labels of Rank are the same as that of H*

    *Rank* ← an *empty matrix*
    **for** *i* from 1 to *row_count(B)* **do**
        *r* = *rank(B[i,])*
        *Rank* [*i*,] = *r*
        *rowname (Rank)* = $S_c$
        *colname (Rank)* = *R*

    **return** (*Rank*)

*Comment: Function to active a subcategory belonging to an activated category based on the items in basket B.*

**function** *subcategory_activation* (*c, B, n, θ, active_subcat*)
    *rank_matrix* = *item_rank(c)*
    *i* ← 0
    *j* ← 0
    **for** *s* in $S_c$ **do**
        *i* ← *i* + 1
        **for** item in *B* **do**
            *j* ← *j* + 1
            *score* ← *score* + $\frac{1}{\sqrt[n]{rank\_matrix[i,\ j]}}$
        **if** *score* > *θ* **then**
            *active_subcat* ← add *s* to the list *active_subcat*
    **return** (*active_subcat*)

*Comment: Process to activate a subcategory. When a new item is added to the basket, the score is updated, and the algorithm checks if any subcategory can be activated.*

*active_subcat* ← an empty array

**while** an *item* is added in *B* **do**
    **for** *c* in *active_cat* **do**
        *a* ← *subcategory_activation* (*c,B,n,θ,active_subcat*)
        *active_subcat* ← add *subcategories* in *a* to the *active_subcat* array

*C. Item Recommendation*

    Only the dishes that belong to the activated subcategories can be recommended. The dishes that correspond to the top *N* largest values of the basket-to-recipe similarities are recommended. Since the dishes may have item sets of varying length, it would be inappropriate to calculate the similarity as $n(I_{dsc} - B)$. This metric will tend to return the item sets with smaller lengths. We use Jaccard similarity between the item sets $I_{dsc}$ and *B* as a measure of basket-to-recipe similarity. Jaccard similarity between two sets *A* and *B* is calculated as follows.

$$J(A, B) = \frac{n(A \cap B)}{n(A \cup B)} \quad (2)$$

    The value of $J(A, B)$ ranges between 0 and 1. For every dish that belongs to an activated subcategory, we calculate the basket-to-recipe similarity, and the top *n* dishes are recommended to the customer.

    The buyer gets a choice to choose from the list of recommended dishes. If the buyer chooses a dish *d* that belongs to the subcategory *s* in category *c*, the algorithm then calculates and displays the yet-to-be-purchased items to prepare the dish *d*, i.e., the set $I_{dsc} - B$, to the buyers and let the buyers choose from the list. The chosen items are added to the basket *B*.

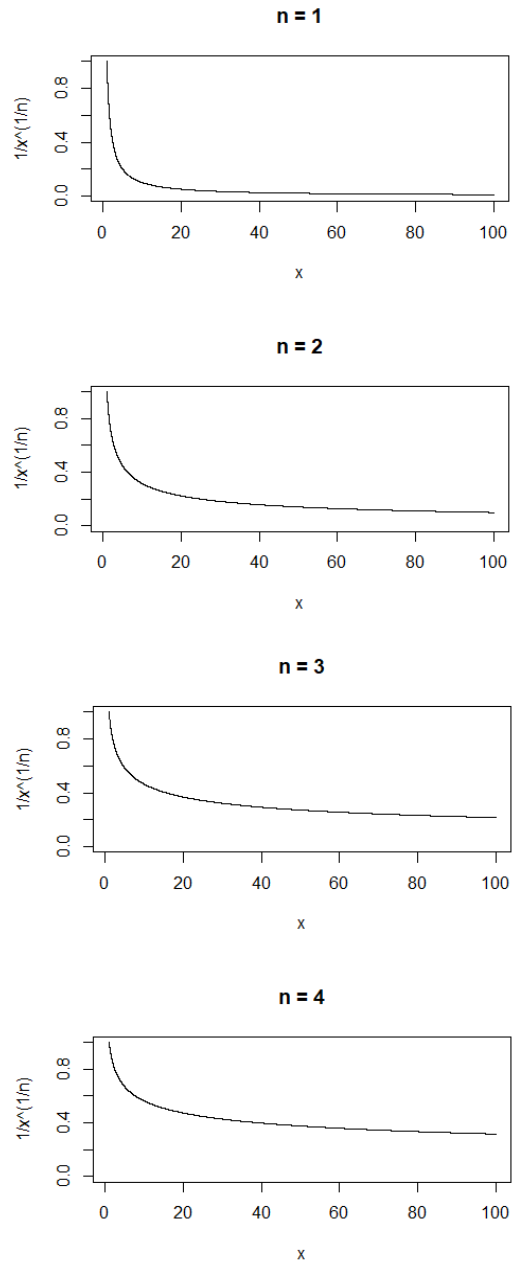

Fig 1. The graphical representation of the function $f(x) = \frac{1}{\sqrt[n]{x}}$ for different values of *n*.

*Algorithm 3. Item recommendation algorithm.*

*active_dish* ← a set containing the *pair* (*dish*, *itemsets*) of the dishes in the *active subcategory*

**function** *recommend_dishes* (*active_dish*)
    *recommend* ← an *empty array*
    *sim* ← an *empty array*
    *i* ← 0
    **for** (*dish*, *itemsets*) in *active_dish* **do**
        *i* ← *i* + 1
        *J* = *Jaccard* (*itemsets*, *B*)
        *sim*[*i*] = (*J*, *dish*)    {storing as pairs}

    *sim* ← **sort**(*sim*) according to decreasing *J*
    *recommend* ← *dish* **for** first *k sim* pairs
    **return** (*recommend*)

**function** *recommend_items* ($I_{dish}$, *B*)
    **return** ($I_{dish} - B$)

IV. RESULTS

For the purpose of our experiment, two categories are selected – *rice* and *chicken*. For each category, we have selected some subcategories. For the category *rice*, we have selected three subcategories, viz. *biryani, fried rice,* and *pulao*. For the category *chicken,* we have selected two subcategories, viz., *Indian* and *Chinese*. For each subcategory, a number of dishes are chosen. For example, few dishes that are selected under the subcategory *pulao* are *peas pulao, tawa pulao, methi pulao, Kashmiri pulao, etc*. A number of recipes for preparing each of these dishes are scraped from the web, and the ingredients (items) used in these recipes are parsed. Finally, a dataset is created with the following fields – (1) food category, (2) food subcategory, (3) dish, (4) items, and (5) recipe.

*A. Data Preprocessing*

For each category, items with the top 5 support values are considered as the food category identifiers. We have found out that the most prominent identifiers for the category *rice* are *long-grain rice, clove, cinnamon, cardamom,* and *ghee,* and for the category *chicken,* are mostly the chicken items like *chicken*, *chicken breast, chicken leg, chicken thigh, and chicken boneless.*

The item with the top global support value is *salt*. It is used in almost all the recipes under our consideration. We ensured that these items, if present, are eliminated from a category identifier list. Activation of a category will automatically activate the subcategory identification algorithms. We choose q = 1 for this experiment, which means that the addition of any one item from a category identifier list will activate that category. and text heads.

*B. Recognizing the category identifiers*

For each category, items with the top 5 support values are considered as the food category identifiers. We have found out that the most prominent identifiers for the category *rice* are *long-grain rice, clove, cinnamon, cardamom,* and *ghee,* and for the category, *chicken* are mostly the chicken items like *chicken*, *chicken breast, chicken leg, chicken thigh, and chicken boneless.*

The item with top global support value is *salt* because it is used in almost all the dishes under our consideration. We ensured that these items, if present, are eliminated from a category identifier list. Activation of a category will automatically activate the subcategory identification algorithms. We choose q = 1 for this experiment, which means that the addition of any one item from a category identifier list will activate that category.

*C. Recognizing the subcategory differentiators*

We have noted that the items with high support values in a given category may not play an important role in differentiating its subcategories. For example, the presence of the item *long-grained rice* in a recipe that belongs to the category *rice* doesn't give us much idea about whether the recipe is for *biryani* or for *fried rice*, although the item has got a very high support value in that category. However, if items like *kewra water*, *mace*, *curd, black peppercorns, and ginger garlic paste* are present in the recipe along with the item *long-grained* rice, then the recipe is more likely to be of *biryani* rather than of *fried rice* or *pulao*.

Table I gives the top 5 subcategory differentiators for each subcategory according to their ranks.

*D. Item recommendation*

With every addition of items in the basket, we have calculated the score corresponding to each activated subcategory using the formula given in (1). Considering a fixed value of the threshold, if we keep increasing the value of n, the score in (1) needs a lesser number of items to cross the threshold. For example, if the threshold ($\theta$) is set to 5 and if we keep adding the items in increasing order of their rank in an empty basket, then it will take 10 items to cross the threshold if the value of *n* is 2 and 83 items with a value of *n* is equal to 1. However, with a value of n more than 6, it stabilizes to a single value.

*E. Item recommendation*

With every addition of items in the basket, we have calculated the score corresponding to each activated subcategory using the formula given in (1). Considering a fixed value of the threshold, if we keep increasing the value of n, the score in (1) needs a lesser number of items to cross the threshold. For example, if the threshold ($\theta$) is set to 5 and if we keep adding the items in increasing order of their rank in an empty basket, then it will take 10 items to cross the threshold if the value of n is 2 and 83 items with a value of n is equal to 1. However, with a value of n more than 6, it stabilizes to a single value.

*F. Item recommendation*

With every addition of items in the basket, we have calculated the score corresponding to each activated subcategory using the formula given in (1). Considering a fixed value of the threshold, if we keep increasing the value of *n*, the score in (1) needs a lesser number of items to cross the threshold. For example, if the threshold ($\theta$) is set to 5 and if we keep adding the items in increasing order of their rank in an empty basket, then it will take 10 items to cross the threshold if the value of n is 2 and 83 items with a value of n is equal to 1. However, with a value of n more than 6, it stabilizes to a single value.

For different values of n and $\theta$ we have calculated the minimum number of items that should be put inside an empty basket in accordance to their increasing value of rank. Table II summarizes that. The values in the cells in the table represent the minimum basket size that will be required for the score in (1) to exceed the corresponding threshold value $\theta$ for a given value of $n$. For example, when the value of $n$ is 3, a minimum of 6 top-ranked items are needed to be put in the empty basket for the score to exceed $\theta = 4$.

From table II it is clear that it doesn't make sense to select a value of $\theta$ as equal to 1. Also, it won't be advisable to choose a very high value of $\theta$, because that will demand a larger basket size for activating any subcategory. It would be better to avoid selecting the value of *n* is equal to 1. And even if we do so, it would be better to limit ourselves to a smaller threshold value. Also, it would be quite useless to select values of n that are greater than 4 or 5 because the cell values tend to settle down to some single value beyond that. The above table can be useful to select a value of n and a value of threshold depending upon the purpose and the problem.

TABLE I. SUBCATEGORY DIFFERENTIATORS FOR EACH SUBCATEGORY ACCORDING TO THEIR RANKS

| Category | Sub-category | Rank |  |  |  |  |
|---|---|---|---|---|---|---|
|  |  | *1* | *2* | *3* | *4* | *5* |
| Rice | *Biryani* | kewra water | mace | curd | black pepper-corn | ginger garlic paste |
| Rice | *Fried Rice* | dark soya sauce | garlic | cabbage | egg | carrot |
| Rice | *Pulao* | cumin seed | almond | cashew nut | green pea | coconut |
| Chicken | *Indian* | cumin | coriander powder | coriander | cilantro | chicken |
| Chicken | *Chinese* | dark soya sauce | corn starch | chicken breast | chicken broth | capsicum |

## V. CONCLUSION

Groceries are often purchased in bulk, and thus a grocer is named after the French word 'grossier', meaning wholesaler. Therefore, a grocery item recommender system, like the one we proposed, should be capable of recommending items in bulk. And so does it. This algorithm can also be viewed as a content-based recommender system that attempts to match the contents of the user's basket with the ingredients contained in different recipes under consideration. However, this algorithm does not suffer from the problems of limited content analysis or over-specialization or the new user problem that a content-based recommender system based on user ratings and item-item content similarity calculations may suffer. The algorithm is also capable of recommending new types of items that the user has not used before. However, it should be noted that this algorithm will only be able to recommend an item if the item is present in the ingredient list of at least one recipe under consideration.

TABLE II. NUMBER OF ITEMS REQUIRED TO EXCEED $\theta$ FOR A GIVEN VALUE OF n WHEN THE ITEMS ARE ADDED IN INCREASING ORDER OF THEIR RANK

|  |  | n |  |  |  |  |  |  |  |  |  |
|---|---|---|---|---|---|---|---|---|---|---|---|
|  |  | *1* | *2* | *3* | *4* | *5* | *6* | *7* | *8* | *9* | *10* |
| Threshold ($\theta$) | *1* | 1 | 1 | 1 | 1 | 1 | 1 | 1 | 1 | 1 | 1 |
|  | *2* | 4 | 3 | 3 | 3 | 3 | 3 | 3 | 3 | 3 | 3 |
|  | *3* | 11 | 5 | 4 | 4 | 4 | 4 | 4 | 4 | 4 | 4 |
|  | *4* | 31 | 7 | 6 | 6 | 5 | 5 | 5 | 5 | 5 | 5 |
|  | *5* | 83 | 10 | 8 | 7 | 7 | 6 | 6 | 6 | 6 | 6 |
|  | *6* | 227 | 14 | 10 | 9 | 8 | 8 | 8 | 7 | 7 | 7 |
|  | *7* | 616 | 18 | 12 | 11 | 10 | 9 | 9 | 9 | 9 | 8 |

ACKNOWLEDGMENT

The first author thanks his friend Soumik Bhusan for introducing this problem. The first author also thanks his brother Rishav Nath and his friend Soaib Khan for their contributions in the data collection process.